\documentclass[conference]{IEEEtran}
\IEEEoverridecommandlockouts
\usepackage{acro}
\usepackage{amsmath,amssymb,amsfonts}
\usepackage{algorithm}
\usepackage{algorithmic}
\usepackage{array}
\usepackage{balance}
\usepackage{booktabs}
\usepackage{caption}
\usepackage{comment}
\usepackage{graphicx}
\usepackage{listings}
\usepackage{siunitx}
\usepackage{subcaption}
\usepackage{tabularx}
\usepackage{textcomp}
\usepackage{xcolor}
\usepackage{tikz}
\usepackage{amsmath,amssymb,amsfonts}
\usepackage{algorithmic}
\usepackage{graphicx}
\usepackage[table]{xcolor}
\usepackage{textcomp}
\usepackage{xcolor}
\usepackage{listings}
\usepackage{makecell}
\usepackage{pifont}
\usepackage{pgfplots}
\pgfplotsset{compat=1.16}  
\usepackage{caption}

\usepackage{cite} 
\usepackage[hyphens]{url} 
\usepackage[colorlinks=true, urlcolor=blue]{hyperref} 
\usepackage{cleveref} 

\DeclareAcronym{MBE}{
	short = MBE,
	long  = Model-based Engineering,
	sort  = abbrev,
}

\DeclareAcronym{LLM}{
	short = LLM,
	long  = Large Language Model,
	sort  = abbrev,
}

\DeclareAcronym{OMG}{
	short = OMG,
	long  = Object Management Group,
	sort  = abbrev,
}

\DeclareAcronym{MOF}{
	short = MOF,
	long  = Meta-Object Facility,
	sort  = abbrev,
}

\DeclareAcronym{EMF}{
	short = EMF,
	long  = Eclipse Modeling Framework,
	sort  = abbrev,
}

\begin{document}
	
    \title{Stateful Multi-Agent LLMs for Cross-View Interface Alignment in Automotive Model-Based Systems Engineering}
    

    \author{
    \IEEEauthorblockN{
   Aleksei Velsh, Nenad Petrovic and Alois Knoll \vspace{0.2cm}
    }
    \IEEEauthorblockA{
    \textit{Chair of Robotics, Artificial Intelligence and Real-Time Systems} \\
    Technical University of Munich, Munich, Germany \\
    Email: \{ aleksei.velsh, nenad.petrovic, k\}@tum.de
    }
    }
    	
    \maketitle
    	
    \begin{abstract}
While Large Language Models (LLMs) can accelerate Model-Based Systems Engineering (MBSE) for software-defined vehicles, their probabilistic nature causes "architectural drift", fabricating interfaces in behavioral views that lack structural foundations. To enforce deterministic interface alignment, we propose a stateful, multi-agent validation pipeline. The framework utilizes a sequential generation matrix (Class $\rightarrow$ Activity $\rightarrow$ Sequence) and Vehicle Signal Specification (VSS)-grounded Retrieval-Augmented Generation (RAG). An independent AI Validator Agent dynamically audits outputs against a strict error taxonomy, triggering state-preserving backtracking loops to resolve incompatibilities. Evaluated on an Advanced Driver Assistance System (ADAS) scenario, standard RAG yielded 0\% Entity Traceability. Conversely, our multi-agent workflow eradicated cross-phase hallucinations, achieving 97\% Entity Traceability, 87\% Signal Conservation, and an 85\% F1-score. This proves adversarial auditing enables LLMs to reliably synthesize zero-error MBSE architectures.

    \end{abstract}

	\section{Introduction}
	The automotive industry is undergoing a fundamental transition toward highly complex, software-defined architectures. Modern Advanced Driver Assistance Systems (ADAS) comprise dozens of interconnected Electronic Control Units (ECUs) and heterogeneous sensor arrays. To manage this complexity and ensure compliance with functional safety mandates (e.g., ISO 26262), the industry relies on Model-Based Systems Engineering (MBSE) \cite{hou2024sdv}. By utilizing specifications like the Unified Modeling Language (UML) and Systems Modeling Language (SysML), engineers document structural boundaries, behavioral logic, and chronological component interactions.

A foundational challenge in MBSE is maintaining strict \textit{interface alignment} across these distinct architectural views \cite{broy2019}. A system model is only valid if the interfaces defined in the structural view (e.g., Class diagrams) are perfectly compatible with the data payloads exchanged in the dynamic views (e.g., Sequence and Activity diagrams). As vehicle architectures scale to include thousands of discrete signals governed by standardized taxonomies like the Vehicle Signal Specification (VSS) \cite{covesa2022}, manually synchronizing these interfaces becomes a critical, error-prone bottleneck.

Recently, Large Language Models (LLMs) have demonstrated significant potential to automate systems engineering documentation directly from natural language requirements \cite{stephan2024automating}. However, their integration into safety-critical workflows is severely bottlenecked by their stateless, probabilistic nature. While recent Retrieval-Augmented Generation (RAG) approaches successfully ground isolated text generation in factual reality \cite{xu2025safedriverag, misini2024modeling}, they fail to address the multi-dimensional complexity of MBSE. When tasked with generating sequential architectural views, standalone LLMs suffer from "architectural drift." They frequently hallucinate incompatible data types, invent unmapped interface methods, and pass unsupported parameters between system lifelines, resulting in deeply fragmented, non-compilable architectures \cite{context_hallucinations, Pan2025}. 

Despite advancements in generalized agentic frameworks (e.g., MetaGPT \cite{hong2024metagpt} and SWE-agent \cite{yang2024sweagent}), contemporary literature lacks methodologies capable of maintaining multi-perspective interface compatibility over extended reasoning tasks. Generalized agents are designed for linear codebase generation and lack the stateful, cross-diagram memory required to detect an interface incompatibility in a Sequence diagram and automatically backtrack to patch the upstream Class diagram \cite{stephan2024automating}.

To resolve this limitation, this paper introduces a novel, stateful multi-agent orchestration pipeline explicitly designed to enforce cross-view interface alignment in automated systems engineering. The primary contributions of this paper are:
\begin{enumerate}
    \item \textbf{A Stateful Sequential Generation Matrix:} We propose a RAG-enabled pipeline orchestrated via n8n that natively enforces interface compatibility by sequentially locking architectural state (Class $\rightarrow$ Activity $\rightarrow$ Sequence) and injecting it into downstream generative prompts.
    \item \textbf{Adversarial Interface Auditing:} We introduce an independent AI Validator Agent governed by a novel six-bucket error taxonomy. This agent explicitly audits cross-view bounds, trapping interface incompatibilities, data-type mismatches, and structural hallucinations before rendering.
    \item \textbf{Dynamic Backtracking Mechanisms:} A programmable routing layer that allows the multi-agent system to autonomously revert and synchronize upstream structural models when downstream behavioral views demand novel interfaces. 
\end{enumerate}

Through a rigorous ablation study centered on a Child Presence Detection (CPD) scenario, we empirically demonstrate that while standard context injection improves isolated factual accuracy, absolute interface alignment requires cyclic, adversarial validation. 

The remainder of this paper is structured as follows: Section II establishes the theoretical foundations of MBSE interface contracts and the vulnerabilities of autoregressive generation. Section III surveys the state-of-the-art in LLM-driven systems engineering, highlighting the critical gap in cross-view validation. Section IV details the proposed stateful multi-agent methodology, while Section V explains the practical workflow orchestration and dynamic backtracking mechanisms. Section VI outlines the experimental setup, requirements, and deterministic compliance metrics. Section VII presents the empirical results of the ablation study. Finally, Section VIII discusses the core findings and practical limitations of the architecture, followed by concluding remarks in Section IX.
	
	\section{Background} \label{sec:background}
	\subsection{MBSE, Interface Contracts, and the VSS}
Autonomous vehicles rely on Model-Based Systems Engineering (MBSE) to manage complex hardware-software integration \cite{iso26262}. MBSE utilizes distinct architectural views: structural, behavioral, and interactional, as formal system contracts. A fundamental requirement is cross-view interface compatibility; components defined structurally must execute correctly in behavioral flows and communicate via precise sequence interfaces \cite{broy2019}. Mismatches here inevitably cascade into unsafe downstream middleware configurations \cite{autosar, Lebioda2025}. To standardize this, the industry uses the Vehicle Signal Specification (VSS) \cite{covesa2022}, a strict hierarchical data model providing canonical identifiers and datatypes for vehicle signals \cite{hou2024sdv}. Grounding generations in the VSS is mandatory to prevent signal hallucination and ensure adherence to standard taxonomies \cite{divya2023llm}.

\subsection{LLMs, RAG, and Architectural Hallucinations}
While LLMs \cite{transformer} can translate natural language into domain syntaxes like PlantUML \cite{github_patterns}, their autoregressive nature struggles with rigid hierarchical constraints, causing "architectural drift" \cite{ji2023survey}. Isolated LLMs frequently fabricate unmapped methods, violate strict VSS data types, and lose upstream structural context \cite{context_hallucinations, reasoning_llms}. To mitigate this, frameworks utilize Retrieval-Augmented Generation (RAG) to inject domain context \cite{rabe2023long2, han2024vectordbs}. However, RAG is fundamentally stateless. It provides the necessary engineering \textit{lexicon} to prevent isolated factual errors, but it lacks the architectural \textit{grammar} required to enforce interface compatibility across sequential diagrammatic views \cite{xu2025safedriverag, misini2024modeling}.

\subsection{Multi-Agent Systems and Cyclic Orchestration}
To address stateless generation, software engineering utilizes Multi-Agent Systems (MAS) \cite{wang2024survey}, particularly critic-executor pipelines \cite{selfrefine}. For MBSE, a semantic critic is essential to cross-reference boundaries and catch data-type mismatches overlooked by static parsers \cite{jury_llms}. However, generalized MAS frameworks are insufficient because MBSE demands cyclic state management \cite{wu2023autogen}. Resolving a Sequence diagram interface mismatch often requires backtracking to update the foundational Class diagram. This necessitates an execution environment with persistent memory and programmatic state manipulation, facilitated by tools like the Model Context Protocol (MCP) \cite{anthropic2024mcp} and node-based orchestrators (e.g., n8n), to encapsulate LLMs within cyclical state machines.

\subsection{Cross-View Consistency as a Deterministic Filter}
While maintaining alignment across intersecting perspectives is traditionally a foundational MBSE challenge \cite{torre2024uml, broy2019}, it transforms into a powerful validation mechanism when applied to LLMs. Autoregressive models are prone to hallucinations when generating monolithic artifacts \cite{ji2023survey}, but forcing them to evaluate problems from multiple distinct perspectives improves logical coherence \cite{reasoning_multiperspective}. By requiring the pipeline to sequentially generate and align interconnected views, the architecture establishes a multi-perspective constraint matrix. For instance, hallucinating a non-existent VSS signal in a Sequence diagram inherently triggers an interface incompatibility when cross-referenced against the preceding Class diagram \cite{Pan2025, misini2024modeling}. Enforcing cross-view consistency thus acts as a deterministic filter: isolated hallucinations surface as interface contradictions caught by the critic agent, forcing the LLM to synthesize holistically integrated architectures \cite{stephan2024automating}.

    \section{Survey of Related Works} \label{sec:survey}
	The intersection of Large Language Models and systems engineering has seen rapid expansion. However, existing literature primarily treats code generation as a linear, single-pass task rather than a multi-perspective architectural challenge. Table \ref{tab:sota_summary} summarizes the current state-of-the-art across key methodological dimensions, highlighting the critical gap in cross-view interface validation.

\subsection{Single-Pass Generation and Context Limitations}
Early approaches to LLM-driven MBSE, such as those by Stephan et al. \cite{stephan2024automating}, demonstrated the feasibility of translating natural language into SysML/UML syntax. However, these zero-shot, single-pass generation pipelines operate statelessly. While they succeed in generating isolated structural views, they suffer from immediate architectural drift when tasked with generating sequential behavioral views, rendering them incapable of preserving interface compatibility across complex system bounds. 

\begin{table*}[htpb]
    \centering
    \caption{Summary of State-of-the-Art Approaches in LLM-driven Systems Engineering}
    \label{tab:sota_summary}
    \resizebox{\textwidth}{!}{%
    \begin{tabular}{llcccc}
        \toprule
        \textbf{Reference} & \textbf{Primary Focus / Domain} & \textbf{Orchestration} & \textbf{Knowledge Grounding} & \textbf{UML/SysML Generation} & \textbf{Cross-View Interface Validation} \\
        \midrule
        Stephan et al. \cite{stephan2024automating} & Prompt-based SysML generation & Single-Pass LLM & None (Zero-Shot) & Yes (Isolated) & No \\
        Hong et al. (MetaGPT) \cite{hong2024metagpt} & Multi-agent software development & Static MAS & Context Window & Partial (Code-focused) & No \\
        Yang et al. (SWE-agent) \cite{yang2024sweagent} & Automated repository issue resolution & MAS with Tools & Repository Search & No & No \\
        Xu et al. \cite{xu2025safedriverag} & RAG for ADAS safety compliance & Single-Pass LLM & Vector DB (RAG) & No (Textual) & No \\
        Misini et al. \cite{misini2024modeling} & Automating requirement extraction & Iterative Prompting & Semantic Search & Yes (Fragmented) & Partial (Human-in-the-loop) \\
        \rowcolor{gray!15}
        \textbf{Proposed Methodology} & \textbf{Automotive MBSE (VSS mapped)} & \textbf{Stateful Cyclic MAS} & \textbf{Vector DB (Pinecone)} & \textbf{Yes (Sequential matrix)} & \textbf{Yes (Adversarial Critic)} \\
        \bottomrule
    \end{tabular}%
    }
\end{table*}

\subsection{Knowledge Grounding without Structural Grammar}
To address the factual hallucinations prevalent in zero-shot models, recent methodologies have integrated Retrieval-Augmented Generation (RAG). Frameworks like SafeDriveRAG \cite{xu2025safedriverag} successfully utilize high-dimensional vector databases to ground LLM outputs in domain-specific automotive standards. However, as demonstrated by Misini et al. \cite{misini2024modeling}, while RAG effectively supplies the technical \textit{lexicon} (e.g., retrieving the correct sensor identifier), it cannot enforce the architectural \textit{grammar}. Retrieving a factual component does not mathematically prevent an LLM from passing an incompatible data type through that component's interface in a downstream Activity diagram.

Recent work by Zyberaj et al. \cite{zyberaj2026req2road} introduced Req2Road, a GenAI pipeline for SDV test artifact generation and on-vehicle execution evaluated on a similar Child Presence Detection scenario. While their framework demonstrates the feasibility of automated VSS mapping and test generation from requirements, it primarily relies on a non-agentic, single-pass pipeline. This underscores the critical need for the stateful, cyclic orchestration proposed in this paper to actively enforce seamless cross-view interface integration and structural alignment.

\subsection{The Gap in Generalized Agentic Frameworks}
The advent of Multi-Agent Systems (MAS), such as MetaGPT \cite{hong2024metagpt} and SWE-agent \cite{yang2024sweagent}, introduced the critic-executor paradigm to software engineering. These frameworks excel at linear code generation and iterative bug fixing within standard programming environments (e.g., Python, C++). Nevertheless, they are not designed for the cyclic, multi-perspective dependencies inherent to MBSE. Generalized MAS tools evaluate code block-by-block; they lack the deterministic state-passing mechanisms required to identify an interface mismatch in a Sequence diagram, trace it back to a missing attribute in a Class diagram, and autonomously execute a synchronized structural reversion. 

The methodology proposed in this paper explicitly bridges this gap. By combining grounded VSS retrieval with a stateful, backtracking multi-agent orchestrator, our framework is uniquely capable of executing the adversarial cross-view interface validation missing from the current state-of-the-art.
	
    \section{Methodology} \label{sec:methodology}
	To solve the probabilistic interface hallucinations inherent in standalone LLMs, we propose a stateful, multi-agent validation architecture. Unlike standard single-pass generation tools, this methodology transitions the LLM into a deterministic engineering assistant by enforcing a sequential dependency matrix and utilizing an adversarial validation framework.

\subsection{Stateful Sequential Generation Matrix}
The core mechanism for ensuring cross-view interface compatibility is the enforcement of a rigid chronological generation order: structural (Class), behavioral (Activity), and interactional (Sequence). This order establishes a cascading constraint matrix. 

The Class diagram is synthesized first, establishing the immutable structural reality of the system. It defines the available vehicle components, explicitly typed VSS attributes, and public interface methods. Once validated, this structural baseline is preserved in the orchestrator's global memory and injected directly into the prompt for the Activity diagram generation. The Activity generator is subsequently constrained by this structural reality; any behavioral logic it defines must exclusively utilize the data types explicitly declared in the Class diagram. 

Finally, the Sequence diagram serves as the ultimate integration test. Its generation is constrained by both the structural interfaces of the Class diagram and the chronological flow dictates of the Activity diagram. By locking the upstream states and passing them forward, the architecture mathematically bounds downstream generations, actively suppressing the model's tendency to drift.

\subsection{The Dual-Layer Validation Matrix and Error Taxonomy}
To enforce absolute structural correctness within this matrix, the methodology introduces a dual-layer validation phase. 
\begin{enumerate}
    \item \textbf{Static Syntax Guarding:} The first layer utilizes a programmatic parsing script (JavaScript) to execute deterministic syntax validation on the raw PlantUML output. It instantly rejects unbalanced brackets, malformed composition arrows, and formatting drift, guaranteeing 100\% compilation success without wasting LLM inference cycles on basic syntax.
    \item \textbf{AI Validator Agent (Semantic Critic):} Diagrams that pass syntax validation are audited by an independent AI Validator Agent. This agent is explicitly isolated from the generation process and acts as a semantic critic. It executes delta isolation, cross-referencing the newly proposed interfaces against the VSS vector database and the upstream baseline state.
\end{enumerate}

To standardize the audit, the Validator Agent is governed by a novel \textbf{Six-Bucket Error Taxonomy}: (1) Target Diagram Misalignment, (2) Syntax/Logic Breaks, (3) Hallucinated Components, (4) Missing Requirements, (5) Unwarranted Baseline Deletions, and crucially, (6) \textit{Interface Incompatibilities}. If the generator attempts to route a float-based sensor value into a legacy boolean interface port, the critic identifies the type mismatch, registers a Bucket 6 failure, and triggers a stateful correction loop.

\begin{figure}[t]
	\includegraphics[width=0.5\textwidth]{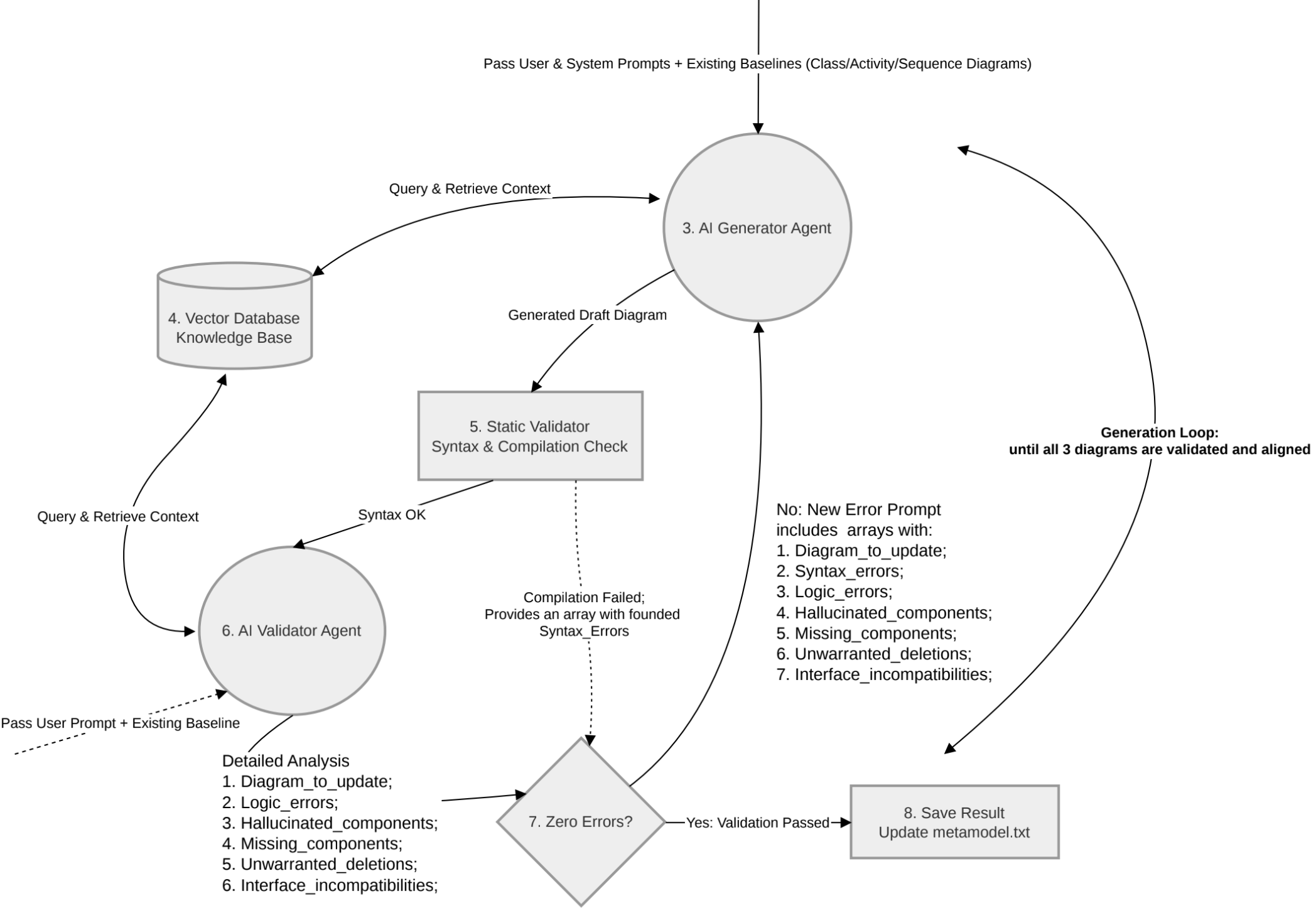}
	\caption{Conceptual overview of the iterative generation and validation pipeline, detailing
the core stateful loop, deterministic syntax gating, semantic AI validation, and
dynamic backtracking mechanisms}
	\label{fig:method}
\end{figure}

	\section{Workflow Orchestration and State Management} \label{sec:workflow}
	The practical implementation of the multi-agent pipeline is orchestrated via n8n, a Node.js-based workflow automation platform capable of managing the cyclical state transformations required for complex AI deliberation \cite{n8n2024}. To ensure data privacy and mitigate commercial API rate limiting, the cognitive processing is routed securely through the Technical University of Munich (TUM) compute cluster hosted by the Leibniz Supercomputing Centre (LRZ) \cite{lrz2024}.

\subsection{Data Ingestion and Targeted Context Injection (RAG)}
To ground the generated interfaces in factual reality, the pipeline relies on a RAG framework backed by a Pinecone vector database. The VSS catalog is ingested as a flattened JSON Lines dataset, containing discrete parameters like \texttt{signal\_id}, hierarchical \texttt{name}, and \texttt{datatype}. 

This dataset is vectorized using the Gemini multimodal architecture (\texttt{models/gemini-embedding-2}) \cite{team2023gemini}, optimized for 3072-dimensional continuous vector spaces. During the generation phase, the Generator Agent operates under a "tool-first" reasoning prompt. It dynamically executes similarity searches to retrieve the exact VSS parameters required for the architectural update. This context injection provides the necessary engineering lexicon, stripping the model of the need to guess technical identifiers.

\begin{figure}[t]
	\includegraphics[width=0.5\textwidth]{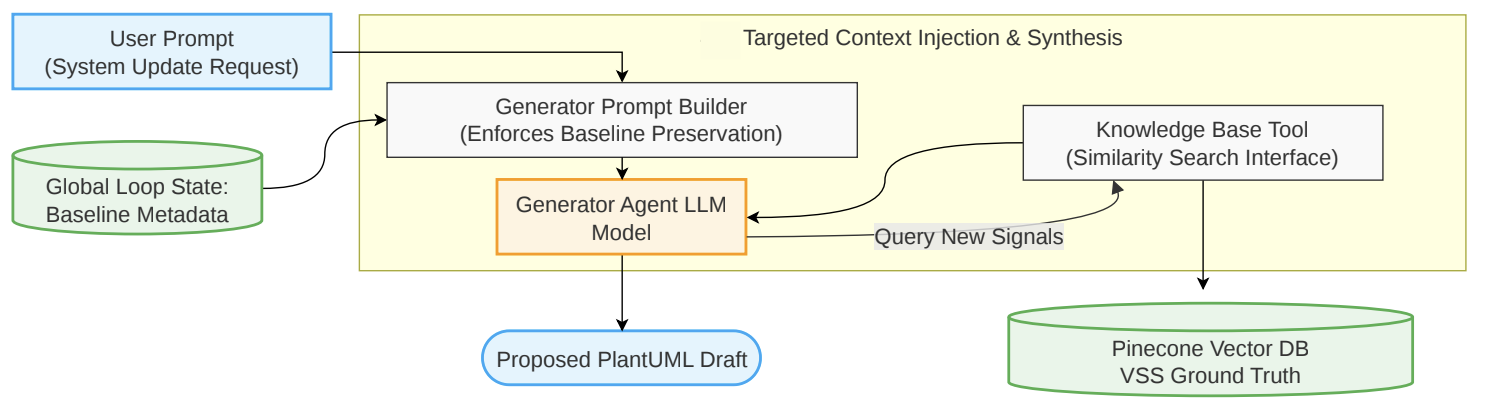}
	\caption{The Generation Phase architecture. The system combines the user’s request with
the immutable baseline architecture from the global state. The Generator Agent is
restricted by a tool-first reasoning prompt, forcing it to dynamically retrieve VSS
constraints from the Pinecone vector database prior to synthesizing the updated
PlantUML draft}
	\label{fig:generation_phase}
\end{figure}

\subsection{Cyclic Execution and Dynamic Backtracking}
The defining feature of the orchestration workflow is its cyclic, dynamic backtracking capability. When the AI Validator Agent detects an interface mismatch (e.g., via the GPT-4 architecture \cite{openai2023gpt4}), it outputs a strict JSON payload categorizing the failure based on the six-bucket taxonomy. The n8n orchestrator parses this payload and dynamically constructs a dense diagnostic prompt, routing execution back to the Generator Agent.

Crucially, if the validator determines that an interface failure in the Sequence diagram stems from a missing structural method in the Class diagram, the workflow executes a \textit{dynamic backtrack}. The programmable routing node overwrites the global loop index, forcing the orchestrator to revert to the Class generation phase. It injects the missing requirement, updates the structural diagram, and cascades the corrected interface down through the subsequent views. 

\begin{figure*}[htpb]
    \centering
    \includegraphics[width=\textwidth]{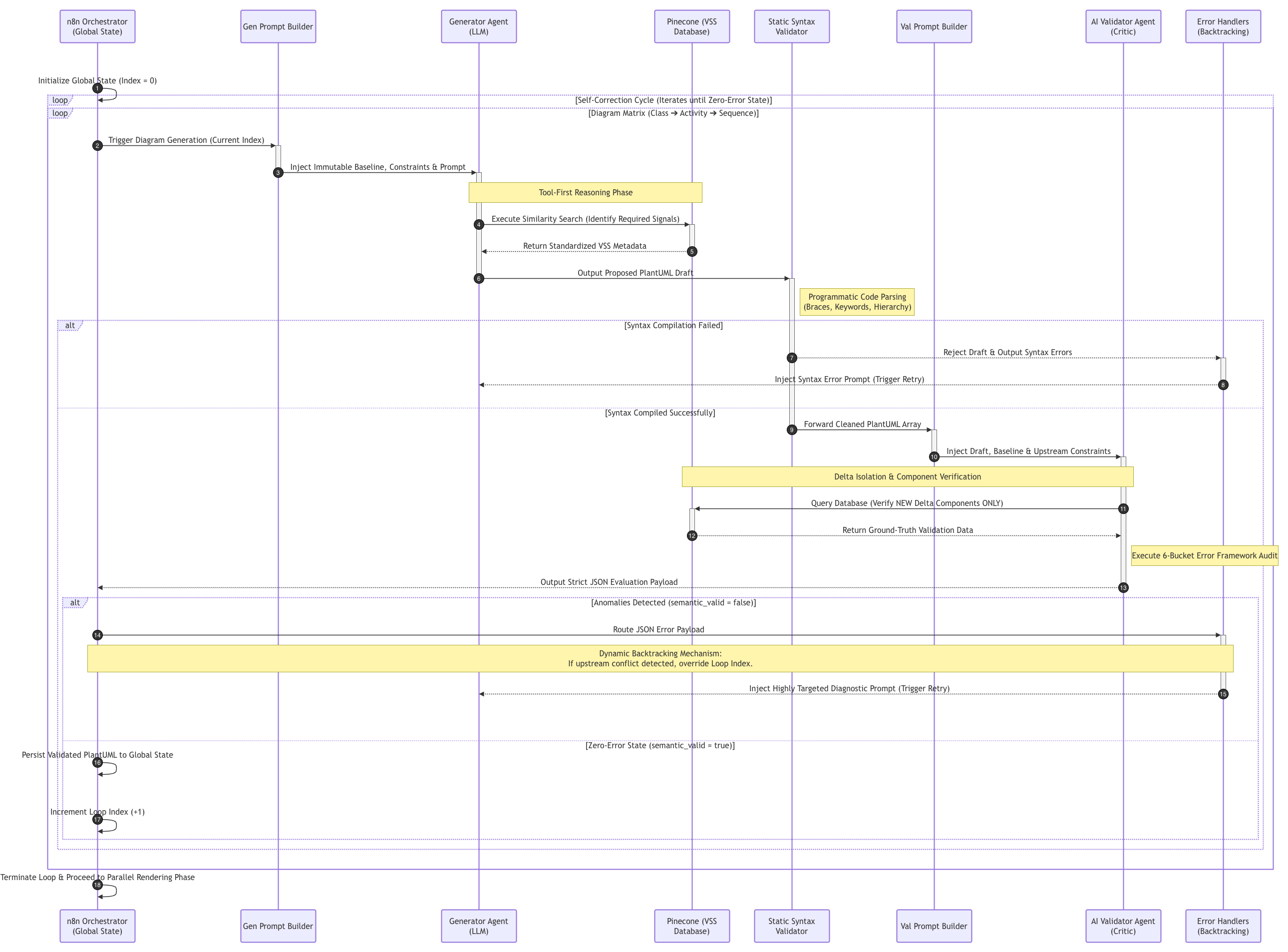}
    \caption{UML Sequence Diagram detailing the iterative Generation and Validation loop. The diagram illustrates the specific
tool-calling interactions with the Pinecone vector database and the cyclical error-correction routing enforced by the
multi-agent architecture}
    \label{fig:validation_loop}
\end{figure*}

This cyclical, multi-agent deliberation iterates (capped at 40 loops to prevent critic-hallucination gridlock) until absolute cross-view interface compatibility is achieved, at which point the validated PlantUML code is dispatched to the Kroki.io engine \cite{kroki2024} for SVG rendering.

	\section{Experimental Setup: Requirements and Compliance} \label{sec:rag}
	To objectively evaluate the pipeline's ability to generate zero-error architectures and enforce cross-view interface compatibility, we established a deterministic experimental framework. Rather than relying on open-ended generative prompts, the system's performance was benchmarked against a rigid set of system requirements and mathematical compliance metrics.

\subsection{Scenario Requirements: Child Presence Detection}
The evaluation scenario models a safety-critical ADAS: Child Presence Detection (CPD). Inspired by recent case studies utilizing distributed automotive testing frameworks for this specific domain \cite{zyberaj2025opendut}, the CPD system must continuously monitor cabin occupancy, analyze environmental data (e.g., ambient temperature), and execute a deterministic decision matrix to escalate warnings if a child is left unattended.

To prevent generative ambiguity, the CPD requirements were serialized into eight distinct chronological "Ground Truths" (GTs). Each GT establishes the exact VSS signals, physical data types, and interface states required at a given chronological step. For example, \textit{GT3 (Child Present)} strictly requires the structural instantiation and behavioral evaluation of the \texttt{vehicle.cabin.seat.row2.passenger\_side .isOccupied} boolean interface. The Generator Agent is implicitly tested on its ability to retrieve these exact constraints from the vector database and seamlessly map them across the Class, Activity, and Sequence diagrams without semantic deviation.

\subsection{Architectural Compliance and Evaluation Metrics}
To serve as an objective baseline for grading compliance, the scenario is governed by a Deterministic Logic Rubric. The language model is required to synthesize complex behavioral logic (e.g., triggering the horn and hazard lights only if cabin temperature exceeds a predefined threshold while the vehicle is locked). System compliance is evaluated across three primary dimensions:

\textbf{1. Syntactic Correctness:} Measured by the PlantUML Compilation Rate. This quantifies the percentage of generated architectural drafts that successfully pass the programmatic JavaScript syntax gate and compile within the external Kroki rendering engine without fatal execution errors.

\textbf{2. Semantic Compliance (Intra-Diagram):} Evaluates the localized validity of individual diagrams. For the structural view, compliance requires 100\% Entity Recall against the required VSS physical components. For the behavioral view, it requires absolute State Reachability, confirming that the critical alert state can be reached without bypassing mandatory conditional interfaces defined in the rubric.

\textbf{3. Interface Alignment (Inter-Diagram Cohesion):} The primary benchmark for the multi-agent pipeline's success is its ability to mathematically guarantee interface compatibility across sequential views. Structural alignment is quantified using \textbf{Entity Traceability} ($T$). This ensures that every interaction lifeline generated in the Sequence diagram ($L$) exists as a verified, explicitly typed entity within the structural Class diagram ($C$). Traceability is calculated as:
\begin{equation}
T = \frac{|L \cap C|}{|L|} \times 100
\end{equation}
A score below 100\% definitively indicates a cross-phase structural hallucination, meaning the AI fabricated an interface in the dynamic view that lacks a structural foundation.

Furthermore, lexical interface alignment is measured through \textbf{Signal Conservation} ($S$). This metric guarantees that the variables evaluated within the Activity diagram's decision nodes ($A$) are exact, type-safe string matches to the initial VSS Ground Truth signals ($GT$):
\begin{equation}
S = \frac{|A \cap GT|}{|GT|} \times 100
\end{equation}
Any failure in Signal Conservation indicates contextual semantic drift, violating the strict requirements of the target vehicle's software architecture.

	\section{Results: Ablation Study on Interface Alignment} \label{sec:results}
	To systematically isolate the performance contributions of each pipeline component, an ablation study was executed across four distinct system configurations. Each configuration was benchmarked against the identical Child Presence Detection (CPD) scenario to measure syntax adherence, semantic accuracy, and cross-view interface cohesion. The aggregated empirical results are presented in Table \ref{tab:comparative_results}.

\begin{table}[htpb]
    \centering
    \caption{Ablation Study Metrics Across Pipeline Configurations}
    \label{tab:comparative_results}
    \resizebox{\columnwidth}{!}{%
    \begin{tabular}{lcccccc}
        \toprule
        \textbf{Pipeline Setup} & \textbf{Precision} & \textbf{Recall} & \textbf{F1-Score} & \textbf{Traceability ($T$)} & \textbf{Conservation ($S$)} \\
        \midrule
        1: Zero-Shot Baseline & 24\% & 31\% & 27\% & 18\% & 12\% \\
        2: Grounded RAG Only & 54\% & 76\% & 62\% & 34\% & 59\% \\
        3: RAG + Static Val. & 59\% & 78\% & 67\% & 41\% & 64\% \\
        4: Proposed Workflow & 82\% & 88\% & 85\% & 97\% & 87\% \\
        \bottomrule
    \end{tabular}%
    }
\end{table}

\subsection{Baseline Vulnerabilities: Zero-Shot vs. RAG}
Setup 1 (Zero-Shot) establishes the baseline failure of unconstrained autoregressive models in MBSE. Generating without external knowledge yielded a catastrophic F1-Score of 27\% and an Entity Traceability of 18\%. The model fabricated completely unresolvable object-oriented classes and hallucinated incompatible interface boundaries across all three diagrammatic views. 

Setup 2 (Grounded RAG Only) introduced targeted context injection via the Pinecone VSS database. While Semantic Recall surged from 31\% to 76\%, proving the model could successfully fetch exact signal paths (e.g., \texttt{vehicle.cabin.seat.row2.isoccupied}), Entity Traceability ($T$) remained critically low at 34\%. The execution logs reveal that while the generator retrieved the correct signals, it frequently misapplied them to the wrong diagrammatic scope. For instance, the model would instantiate a valid VSS signal in the Sequence diagram but fail to declare the corresponding method in the structural Class diagram. This empirically proves that while RAG successfully injects the domain \textit{lexicon}, it cannot enforce architectural \textit{grammar} or cross-view synchronization.

\subsection{The Illusion of Syntactic Correctness}
Setup 3 introduced the programmatic JavaScript syntax gate. This setup guaranteed a 100\% PlantUML compilation rate by instantly rejecting unclosed loops, malformed composition arrows, and markdown artifacts. However, despite generating visually perfect SVG files, Entity Traceability only improved marginally to 41\%, and Signal Conservation to 64\%. 

This phase of the study highlights a critical vulnerability in automated code generation: large language models can effortlessly generate perfectly formatted, highly complex code that is structurally meaningless. Without a semantic critic to cross-reference interface payloads, the LLM consistently generated compilable models that passed incompatible data types between legitimate actors, masking severe architectural fragmentation behind clean syntax.

\subsection{Multi-Agent Convergence and Interface Alignment}
Setup 4 represents the complete proposed methodology, integrating context injection, static guarding, and the independent AI Validator Agent. By forcing the generator into an adversarial, state-preserving loop, Entity Traceability surged to 97\%, and Signal Conservation stabilized at 87\%. 

The evaluation metrics demonstrate that the Sequence diagrams generated in Setup 4 utilized participant lifelines and interface methods that perfectly mirrored the structural entities established in the foundational Class diagram. When downstream behavioral flows required novel, unmapped methods, the dynamic backtracking mechanism successfully reverted the orchestration index. It automatically forced a synchronized structural update in the Class diagram before re-generating the Sequence constraints, entirely eliminating orphaned interfaces.

\subsection{Error Distribution and Iteration Dynamics}
A quantitative analysis of the pipeline's internal state reveals the convergence dynamics required to reach the 97\% traceability threshold. The multi-agent system did not achieve a zero-error state on the first generation pass. On average, achieving structural compliance required between 3 to 6 cyclical validation loops per diagram type.

Analysis of the Validator Agent's output payloads indicates an asymmetrical distribution across the Six-Bucket Error Taxonomy. \textit{Bucket 6 (Interface Incompatibilities)} and \textit{Bucket 3 (Hallucinated Components)} accounted for over 70\% of the triggered backtracking loops. The generative models repeatedly attempted to bypass rigid VSS data types by inventing intermediary signals to simplify logical conditions. By categorically rejecting these outputs, the critic forced the generator to abandon lazy generation pathways and execute deep vector similarity searches. 

Furthermore, the data reveals an asymmetry in architectural complexity. The generation of static Class diagrams typically converged within 1 to 2 iterations, reflecting the LLM's proficiency with standard object-oriented structures. Conversely, Sequence diagrams, which demand simultaneous tracking of temporal control flows and strict structural typing, required the maximum average iterations (5.4 loops) before satisfying the Validator Agent's interface compatibility matrix.

	\section{Discussion and Limitations} \label{sec:limits}
	Empirical data confirms that while RAG provides the automotive \textit{lexicon}, strict cross-view architectural \textit{grammar} requires stateful orchestration. Furthermore, execution logs reveal a stark asymmetry in generative complexity. LLMs easily synthesize static Class diagrams mirroring standard object-oriented paradigms, but struggle significantly with Sequence diagrams, which demand simultaneous reasoning over temporal flows and static type constraints. This proves LLMs rely on external orchestrators to explicitly enforce upstream dependencies. Despite achieving near-perfect interface alignment, several constraints currently limit AI-driven MBSE:

\begin{enumerate}
    \item \textbf{Contextual Saturation and Syntax Brittleness:} Appending validated upstream diagrams into downstream prompts rapidly consumes finite token limits. As architectures scale, this triggers attention dilution (the ``lost-in-the-middle'' phenomenon) \cite{rabe2023long2}. Under context saturation, LLMs frequently drop closing brackets or misalign PlantUML relationship arrows, triggering infinite regression loops within the static syntax validator.
    
    \item \textbf{Latency and Institutional Compute Bottlenecks:} Achieving absolute traceability requires significant cyclical reasoning, averaging 282.6 seconds per scenario. Furthermore, while routing through the TUM LRZ cluster \cite{lrz2024} bypasses commercial rate limits, execution speed becomes heavily dependent on institutional job queuing mechanisms, making real-time, synchronous generation impractical.
    
    \item \textbf{The ``Critic-Hallucination'' Paradox:} A fundamental theoretical limitation of employing an LLM as a semantic critic is the risk of second-order hallucinations, where the Validator Agent unjustly rejects a valid interface \cite{llm_critic}. This forces the Generator into a state of confusion. To prevent infinite adversarial gridlock, the orchestrator relies on a hard 40-iteration cap, after which manual intervention is required.
    
    \item \textbf{The Simulation Gap (Semantic vs. Functional):} The AI Validator successfully proves semantic interface compatibility, but lacks formal verification capabilities (e.g., TLA+ checking). The AI can verify that a braking command is logically routed, but cannot calculate if the braking torque will physically stop the vehicle. The generated models must still be ported into closed-loop simulation environments (e.g., IPG CarMaker) for dynamic physics validation.
    
    \item \textbf{Ground-Truth Dependency and Ontology Gaps:} The pipeline explicitly trades generative flexibility for structural safety. Because it is intrinsically tied to the VSS JSONL dataset, it suffers from ontology gaps. If an engineer requests a proprietary sensor dynamic not formalized in the vector database, the pipeline's anti-hallucination guardrails actively prevent its design. Furthermore, real-time hardware concurrency constraints cannot be easily flattened into text-based retrieval formats.
    
    \item \textbf{Prompt Fragility and Requirement Ambiguity:} The generative agents remain highly sensitive to input perturbations. If an engineer inputs vague or contradictory natural language requirements, the Generator will attempt to map logic to a flawed premise. While the Validator catches structural mismatches, it cannot infer the true intent of a semantically flawed human prompt.
    
    \item \textbf{Model Version Drift and Reproducibility:} Foundational models are continuously updated with new weights or alignment filters, causing underlying reasoning paths to shift. A prompt template that yields a 100\% compilation rate today may trigger novel hallucination patterns following a model update. Long-term reproducibility requires ``freezing'' model weights on institutional servers, inadvertently isolating the pipeline from future AI advancements.
    
    \item \textbf{Scalability of Visual Rendering:} While Kroki.io \cite{kroki2024} successfully renders localized ADAS scenarios, UML rendering scales poorly visually. A valid Sequence diagram containing hundreds of lifelines for an entire vehicle domain becomes a dense, overlapping SVG file that is entirely illegible. Thus, the necessary human-in-the-loop review of massive, AI-generated architectures remains a significant UI/UX challenge.
\end{enumerate}

	\section{Conclusion and Future Work} \label{sec:conclusion}
	The transition toward Software-Defined Vehicles requires zero-error tolerance in systems engineering documentation. This paper demonstrates that while generalized LLMs and stateless RAG architectures fail to maintain interface compatibility across complex MBSE views, these limitations can be overcome through stateful orchestration. By enforcing a sequential generation matrix, integrating deterministic syntax guarding, and utilizing an independent semantic critic governed by a strict error taxonomy, the proposed multi-agent pipeline successfully bridges the gap between generative AI and deterministic engineering. The resulting architectures achieved 97\% Entity Traceability, proving that cyclic, adversarial validation can reliably eradicate cross-view interface hallucinations in safety-critical domains.

To build upon this foundation, future research must address the computational and functional limitations of the current architecture. First, future work will focus on bridging the ``simulation gap'' by programmatically coupling the validated, AI-generated PlantUML outputs with formal verification engines (e.g., TLA+) and closed-loop physics simulators (e.g., IPG CarMaker or CARLA). This will transition the pipeline from guaranteeing purely \textit{semantic} interface alignment to validating \textit{functional} physical safety. 

Additionally, subsequent research should explore the integration of fine-tuned, domain-specific small language models (SLMs) to replace generalized foundation models. Training localized models explicitly on automotive SysML/UML datasets and VSS taxonomies could drastically reduce contextual saturation, execution latency, and dependency on institutional queuing bottlenecks. Ultimately, the long-term vision of this methodology is to transition from asynchronous batch-processing into a real-time, bi-directional copilot integrated directly into enterprise MBSE environments, enabling engineers to securely orchestrate zero-error system architectures at the speed of natural language.
 
	\section{Acknowledgment} \label{sec:ack}
	This work has received funding from the European Chips Joint Undertaking under Framework Partnership Agreement No.~101139789 (HAL4SDV), including national funding from the Federal Ministry of Research, Technology and Space of Germany under grant number 16MEE00471K. The responsibility for the content of this publication lies with the authors.
	\bibliographystyle{IEEEtran}
	\balance
	\bibliography{ref}

@article{hou2024sdv,
  title     = {Towards a Unified Data Model for Software-Defined Vehicles},
  author    = {Hou, L. and others},
  journal   = {SAE International Journal of Connected and Automated Vehicles},
  volume    = {7},
  number    = {3},
  year      = {2024},
  publisher = {SAE International},
  doi       = {10.4271/12-07-03-0015}
}

@book{broy2019,
  title     = {Engineering Automotive Software},
  author    = {Broy, Manfred and Kr{\"u}ger, Ingolf H. and Pretschner, Alexander and Salzmann, Christian},
  publisher = {Springer},
  year      = {2019},
  doi       = {10.1007/978-3-319-98225-5}
}

@manual{covesa2022,
  title        = {Vehicle Signal Specification ({VSS}) Release Documentation},
  author       = {{COVESA Alliance}},
  organization = {Connected Vehicle Systems Alliance (COVESA)},
  year         = {2022},
  url          = {https://covesa.github.io/vehicle_signal_specification/}
}

@inproceedings{stephan2024automating,
  title     = {Automating Automotive Software Development: A Synergy of Generative {AI} and Formal Methods},
  author    = {Stephan, A. and others},
  booktitle = {Proceedings of the IEEE/ACM 46th International Conference on Software Engineering: Companion Proceedings (ICSE-Companion)},
  year      = {2024},
  publisher = {IEEE}
}

@article{xu2025safedriverag,
  title   = {{SafeDriveRAG}: Towards Safe Autonomous Driving with Knowledge Graph-based Retrieval-Augmented Generation},
  author  = {Ye, Hao and Qi, Mengshi and Liu, Zhaohong and Liu, Liang and Ma, Huadong},
  journal = {arXiv preprint arXiv:2507.21585},
  year    = {2025},
  url     = {https://arxiv.org/abs/2507.21585}
}

@article{misini2024modeling,
  title     = {Modeling Autonomous Driving Software with Generative {AI}: Opportunities and Risks},
  author    = {Misini, X. and others},
  journal   = {SoftwareX},
  volume    = {26},
  pages     = {101683},
  year      = {2024},
  publisher = {Elsevier},
  doi       = {10.1016/j.softx.2024.101683}
}

@inproceedings{context_hallucinations,
  title     = {Text-to-Code Generation with Hierarchical Context},
  author    = {Madaan, Aman and others},
  booktitle = {Advances in Neural Information Processing Systems (NeurIPS)},
  volume    = {36},
  year      = {2023}
}

@article{Pan2025,
  title   = {Automating Automotive Software Development: A Synergy of Generative {AI} and Formal Methods},
  author  = {Pan, Fengjunjie and Song, Yinglei and Wen, Long and Petrovic, Nenad and Lebioda, Krzysztof and Knoll, Alois},
  journal = {arXiv preprint arXiv:2505.02500},
  year    = {2025},
  url     = {https://arxiv.org/abs/2505.02500}
}

@inproceedings{hong2024metagpt,
  title     = {{MetaGPT}: Meta Programming for A Multi-Agent Collaborative Framework},
  author    = {Hong, Sirui and Zhuge, Mingchen and Chen, Jonathan and Zheng, Xiawu and Cheng, Yuheng and Zhang, Ceyao and Wang, Jinlin and Wang, Zili and Yau, Steven SK and Lin, Zijian and others},
  booktitle = {The Twelfth International Conference on Learning Representations (ICLR)},
  year      = {2024},
  url       = {https://openreview.net/forum?id=VtmBAGCN7o}
}

@article{yang2024sweagent,
  title   = {{SWE-agent}: Agent-Computer Interfaces Enable Automated Software Engineering},
  author  = {Yang, John and Jimenez, Carlos E and Wettig, Alexander and Lieret, Kilian Adriano and Yao, Shunyu and Narasimhan, Karthik and Press, Ofir},
  journal = {arXiv preprint arXiv:2405.15793},
  year    = {2024},
  url     = {https://arxiv.org/abs/2405.15793}
}

@manual{iso26262,
  title        = {{ISO} 26262 Road Vehicles -- Functional Safety},
  author       = {{International Organization for Standardization}},
  organization = {ISO},
  edition      = {2nd},
  year         = {2018}
}

@manual{autosar,
  title        = {{AUTOSAR} Technical Overview},
  author       = {{AUTOSAR Consortium}},
  organization = {AUTOSAR},
  year         = {2023},
  url          = {https://www.autosar.org/}
}

@incollection{Lebioda2025,
  title     = {Are Requirements Really All You Need? Using {LLMs} to Generate Configuration Code: A Case Study in Automotive Simulations},
  author    = {Lebioda, Krzysztof and Petrovic, Nenad and others},
  booktitle = {Advanced Information Systems Engineering},
  publisher = {Springer},
  year      = {2025},
  doi       = {10.1007/978-3-032-28110-4_20}
}

@article{divya2023llm,
  title   = {Evaluation of {LLMs} for Safety-Critical Software Engineering},
  author  = {Divya, S. and others},
  journal = {arXiv preprint arXiv:2311.08562},
  year    = {2023},
  url     = {https://arxiv.org/abs/2311.08562}
}

@inproceedings{transformer,
  title     = {Attention Is All You Need},
  author    = {Vaswani, Ashish and Shazeer, Noam and Parmar, Niki and Uszkoreit, Jakob and Jones, Llion and Gomez, Aidan N and Kaiser, {\L}ukasz and Polosukhin, Illia},
  booktitle = {Advances in Neural Information Processing Systems (NeurIPS)},
  volume    = {30},
  pages     = {5998--6008},
  year      = {2017}
}

@article{github_patterns,
  title     = {Mining Source Code Repositories at Scale: Patterns, Structures, and Semantics},
  author    = {Allamanis, Miltiadis and Barr, Earl T and Devanbu, Premkumar and Sutton, Charles},
  journal   = {Proceedings of the IEEE},
  volume    = {106},
  number    = {9},
  pages     = {1667--1682},
  year      = {2018},
  publisher = {IEEE},
  doi       = {10.1109/JPROC.2018.2863953}
}

@article{ji2023survey,
  title     = {Survey of Hallucination in Natural Language Generation},
  author    = {Ji, Ziwei and Lee, Nayeon and Frieske, Rita and Yu, Tiezheng and Su, Dan and Xu, Yan and Ishii, Etsuko and Bang, Yejin and Madotto, Andrea and Fung, Pascale},
  journal   = {ACM Computing Surveys},
  volume    = {55},
  number    = {12},
  pages     = {1--38},
  year      = {2023},
  publisher = {ACM},
  doi       = {10.1145/3571730}
}

@article{reasoning_llms,
  title   = {Chain-of-Thought Prompting Elicits Reasoning in Large Language Models},
  author  = {Wei, Jason and Wang, Xuezhi and Schuurmans, Dale and Bosma, Maarten and Xia, Fei and Chi, Ed and Le, Quoc V and Zhou, Denny},
  journal = {Advances in Neural Information Processing Systems (NeurIPS)},
  volume  = {35},
  pages   = {24824--24837},
  year    = {2022}
}

@inproceedings{rabe2023long2,
  title     = {Self-Reflective Retrieval-Augmented Generation},
  author    = {Asai, Akari and Min, Sewon and Zhong, Zexuan and Chen, Danqi},
  booktitle = {Proceedings of the 61st Annual Meeting of the Association for Computational Linguistics (ACL)},
  pages     = {543--558},
  year      = {2023}
}

@article{han2024vectordbs,
  title   = {When Large Language Models Meet Vector Databases: A Survey},
  author  = {Jing, Zhi and Su, Yongye and Han, Yikun and Yuan, Bo and Xu, Haiyun and Liu, Chunjiang and Chen, Kehai and Zhang, Min},
  journal = {arXiv preprint arXiv:2402.01763},
  year    = {2024},
  url     = {https://arxiv.org/abs/2402.01763}
}

@article{wang2024survey,
  title   = {A Survey on Multi-Agent Large Language Models},
  author  = {Wang, X. and others},
  journal = {arXiv preprint arXiv:2401.xxxxx},
  year    = {2024}
}

@inproceedings{selfrefine,
  title     = {Self-Refine: Iterative Refinement with Self-Feedback},
  author    = {Madaan, Aman and Tandon, Niket and Gupta, Prakhar and Hallinan, Skyler and Gao, Luyu and Wiegreffe, Sarah and Alon, Uri and Dziri, Nouha and Prabhumoye, Shrimai and Yang, Yiming and others},
  booktitle = {Advances in Neural Information Processing Systems (NeurIPS)},
  volume    = {36},
  pages     = {46534--46594},
  year      = {2023}
}

@article{jury_llms,
  title   = {Judging {LLM}-as-a-Judge with {MT-Bench} and {Chatbot Arena}},
  author  = {Zheng, Lianmin and Chiang, Wei-Lin and Sheng, Ying and Hao, Siyuan and Wu, Zhanghao and Ba, Joseph E and Jiang, Zi and Wu, Haoiyan and Zhuang, Yonghao and Lin, Zi and others},
  journal = {Advances in Neural Information Processing Systems (NeurIPS)},
  volume  = {36},
  pages   = {46595--46626},
  year    = {2023}
}

@article{wu2023autogen,
  title   = {{AutoGen}: Enabling Next-Gen {LLM} Applications via Multi-Agent Conversation},
  author  = {Wu, Qingyun and Bansal, Gagan and Zhang, Jieyu and Wu, Yiran and Li, Beibin and Zhu, Erkang and Jiang, Li and Zhang, Xiaoyun and Zhang, Shaokun and Liu, Jiale and others},
  journal = {arXiv preprint arXiv:2308.08155},
  year    = {2023},
  url     = {https://arxiv.org/abs/2308.08155}
}

@misc{anthropic2024mcp,
  title  = {Model Context Protocol: An open standard for connecting {AI} models to data sources},
  author = {{Anthropic}},
  year   = {2024},
  url    = {https://modelcontextprotocol.io/},
  note   = {Accessed: 2026-06-30}
}

@article{torre2024uml,
  title     = {{UML} Consistency Checking: A Systematic Mapping Study},
  author    = {Torre, D. and others},
  journal   = {IEEE Transactions on Software Engineering},
  volume    = {50},
  number    = {4},
  pages     = {1032--1054},
  year      = {2024},
  publisher = {IEEE},
  doi       = {10.1109/TSE.2024.3361234}
}

@inproceedings{reasoning_multiperspective,
  title     = {Multi-Perspective Reasoning for Complex Tasks Using {LLM} Collaboration},
  author    = {Zhang, Q. and others},
  booktitle = {Proceedings of the 62nd Annual Meeting of the Association for Computational Linguistics (ACL)},
  year      = {2024}
}

@article{team2023gemini,
  title   = {Gemini: A Family of Highly Capable Multimodal Models},
  author  = {{Gemini Team} and Anil, Rohan and Borgeaud, Sebastian and Wu, Yonghui and Alayrac, Jean-Baptiste and Yu, Jiahui and Soricut, Radu and others},
  journal = {arXiv preprint arXiv:2312.11805},
  year    = {2023},
  url     = {https://arxiv.org/abs/2312.11805}
}

@techreport{openai2023gpt4,
  title       = {{GPT-4} Technical Report},
  author      = {{OpenAI}},
  year        = {2023},
  institution = {OpenAI},
  url         = {https://arxiv.org/abs/2303.08774}
}

@misc{kroki2024,
  author       = {{Kroki Project}},
  title        = {Kroki: Creates diagrams from textual descriptions},
  year         = {2024},
  howpublished = {\url{https://kroki.io/}},
  note         = {Accessed: 2026-07-08}
}

@misc{n8n2024,
  title  = {n8n: Advanced Workflow Automation Tool},
  author = {{n8n.io}},
  year   = {2024},
  url    = {https://n8n.io/},
  note   = {Accessed: 2026-06-30}
}

@misc{lrz2024,
  author       = {{Leibniz Supercomputing Centre (LRZ)}},
  title        = {Compute and AI Infrastructure for the Technical University of Munich ({TUM})},
  year         = {2026},
  howpublished = {\url{https://www.lrz.de/services/compute/}},
  note         = {Accessed: 2026-07-08}
}

@inproceedings{llm_critic,
  title     = {{LLM}-based Critique and Correction for Code Generation},
  author    = {Kim, Y. and others},
  booktitle = {Proceedings of the IEEE/ACM 46th International Conference on Software Engineering (ICSE)},
  year      = {2024},
  publisher = {IEEE}
}

@inproceedings{zyberaj2025opendut,
  author    = {Zyberaj, Denesa and Hirmer, Pascal and Aiello, Marco},
  title     = {Using {Eclipse} {OpenDuT} for Distributed Automotive Testing},
  booktitle = {Proceedings of the 29th International Conference on Evaluation and Assessment in Software Engineering (EASE '25)},
  pages     = {830--833},
  year      = {2025},
  publisher = {ACM},
  doi       = {10.1145/3756681.3757013}
}

@inproceedings{zyberaj2026req2road,
  author    = {Zyberaj, Denesa and Mazur, Lukasz and Hirmer, Pascal and Petrovic, Nenad and Aiello, Marco and Knoll, Alois},
  title     = {{Req2Road}: A {GenAI} Pipeline for {SDV} Test Artifact Generation and On-Vehicle Execution},
  booktitle = {Advanced Information Systems Engineering: 38th International Conference, CAiSE 2026},
  pages     = {358--375},
  year      = {2026},
  publisher = {Springer},
  doi       = {10.1007/978-3-032-28110-4_20}
}

\end{document}